\documentclass[aps,prl,twocolumn,superscriptaddress,floatfix,preprintnumbers,showpacs]{revtex4-1}
\usepackage[T1]{fontenc}
\usepackage[latin9]{inputenc}
\usepackage{babel}
\usepackage{verbatim}
\usepackage{amsmath}
\usepackage{amssymb}
\usepackage{stmaryrd}
\usepackage{xcolor}
\usepackage{graphicx,graphics}
\usepackage[unicode=true,pdfusetitle,
 bookmarks=true,bookmarksnumbered=false,bookmarksopen=false,
 breaklinks=false,pdfborder={0 0 1},backref=false,colorlinks=false]
 {hyperref}
\usepackage{tikz}
\usetikzlibrary{positioning}
\usetikzlibrary{calc}
\usepackage{calc}
\usetikzlibrary{arrows, decorations.markings}

\newcommand{\bx}{\textbf{x}}


\begin{document}
\title{KPZ physics and phase transition in a classical single random walker under continuous
measurement}

\author{Tony Jin}
\email{zizhuo.jin@unige.ch}
\affiliation{DQMP, University of Geneva, 24 Quai Ernest-Ansermet, CH-1211 Geneva, Switzerland}
\author{David G. Martin}
\email{dgmartin@uchicago.edu}
\affiliation{Enrico Fermi Institute, The University of Chicago, 933 E 56th St, Chicago, Illinois 60637, USA}

\begin{abstract}
We introduce and study a new model consisting of a single classical random
walker undergoing continuous monitoring at rate $\gamma$ on a discrete lattice. Although such a continuous measurement cannot affect physical observables, it has a non-trivial effect on the probability distribution of the random walker. 
At small $\gamma$, we show analytically that the time-evolution of the latter can be mapped to the Stochastic Heat Equation (SHE).
In this limit, the width of the log probability thus follows a Family-Vicsek scaling law, $N^{\alpha}f(t/N^{\alpha/\beta})$, with roughness and growth exponents corresponding to the Kardar-Parisi-Zhang (KPZ) universality class, i.e $\alpha^{\rm{1D}}_{\rm{KPZ}}=1/2$ and $\beta^{\rm{1D}}_{\rm{KPZ}}=1/3$ respectively.
When $\gamma$ is increased outside this regime, we find numerically in 1D a crossover from the KPZ class to a new universality class characterized by exponents $\alpha^{1\rm{D}}_{\text{M}}\approx 1$ and $\beta^{1\rm{D}}_{\text{M}}\approx 1.4$. In 3D, varying $\gamma$ beyond a critical value $\gamma^c_{\rm{M}}$ leads to a phase transition from a smooth phase that we identify as the Edwards-Wilkinson (EW) class to a new universality class with $\alpha^{3\rm{D}}_{\text{M}}\approx1$.
\end{abstract}
\maketitle
Universality is a pillar concept of statistical physics, classical
and quantum alike. The fact that, under renormalization, different microscopic models can
lead to the same scale invariant theory has been the key idea for understanding second-order phase transitions. 
In particular, the concept of universality classes has found an extremely fertile ground within the study of dynamical interfaces for which a scale invariance property has been reported and documented \citep{FractalconceptsBarbarasi,Geza_ReviewUniversalityclassesnonequilibrium}.
In this context, one particular fixed point has attracted
a tremendous interest in the previous decades: the Kardar-Parisi-Zhang
(KPZ) universality class and its iconic $1/3$ growth exponent \citep{KPZpaper,ReviewKPZ} in 1D. Beyond the eponym KPZ equation, it has been found in a variety of models describing
growing interfaces such as the ballistic
deposition model \citep{Ballisticdeposition}, the Eden model \citep{Eden1,Eden2},
or the restricted solid-on-solid model \citep{RSOS}. 
Perhaps more surprisingly, in the recent years, it has also been discovered in a variety of quantum phenomena such as the growth of entanglement entropy in
random unitary circuits \citep{EntanglementRUCNahum}, stochastic
conformal field theory \citep{BernardLeDoussalStochasticCFT}, noisy
fermions \citep{QKPZ} and transport properties of dipolar spin ensembles
\citep{KPZDiamond} and integrable spin chains \citep{DeNardisMedenjak_UniversalitySpintransport,SuperSuperDiffusionDeNardis,BlocKPZspinchains,JBlochKPZ}. 

Continuous or weak measurement has enjoyed considerable
interest in the previous decades within the quantum community as it provides a non-destructive way to obtain information about a given quantum system
\citep{Aharanovweakmeasurement,ReviewHaroche}.
Its advent led to many interesting applications such as quantum Zeno effects \citep{QZenoeffectItano},
quantum trajectories \citep{WeakmeasurementDoubleslit}, quantum Maxwell
demons \citep{QMaxwellDemon}, or direct observation of quantum jumps
\citep{catchreverseQjumps}. Recently, a number of studies investigated the consequences
of repeated projections or continuous monitoring on the evolution
of quantum many-body systems. For systems undergoing both a random unitary evolution and measurements, a result that has aroused considerable
interests lately is the existence of a Measurement-Induced Phase Transition (MIPT) in the entanglement entropy \citep{caoEntanglementFermionChain2019,NahumMeasurementinducedtransition2019,MeasurementinducedFisher,MeasurementinducedSchomerus,MeasurementinducedAltman,HuseMeasurementinduced,HuseMeasurementinduced3,HuseMeasurementinduced2,LavasaniMeasurementinduced,buchholdEffectiveTheoryMeasurementInduced2021,Xhek1,Xhek2,Altman_PowerLawNegativity}.
Most of these contributions focus on entanglement or R\'{e}nyi entropies, i.e information-related  quantities which are likely salients in classical systems as well.
As such, it is natural to wonder whether the same phenomenology of MIPT also features in classical physics. 

In this paper we unveil a connection between KPZ physics and \emph{classical} information theory by studying a single classical random walker undergoing continuous
monitoring and, relying on this connection, we show that this system presents a MIPT in 3D.  

We first present the framework that we use to model weak, continuous measurements on a generic Markov process.
We then focus on the specific case of a single random walker diffusing on a lattice with the occupancy at each site being continuously monitored.

When the measurement rate $\gamma$ is small, we find in 1D that the standard deviation of the log probability follows a Family-Vicsek scaling law
with roughness and growth exponents corresponding to the KPZ universality class, i.e $\alpha_{\text{KPZ}}^{\rm{1D}}=1/2$ and $\beta_{\text{KPZ}}^{\rm{1D}}=1/3$ respectively \citep{KPZpaper}.
By performing a perturbative analysis around $\gamma=0$, we show analytically that this KPZ-like behavior is due to a direct mapping of the dynamics onto the Stochastic Heat Equation (SHE). As $\gamma$ is increased further, we see numerically in 1D a size-dependent crossover between the KPZ regime and a new universality class characterized by different exponents $\alpha^{1\rm{D}}_{\text{M}}\approx1$ and $\beta^{1\rm{D}}_{\text{M}} \approx 1.4$. In 3D, instead of a crossover, we see a phase transition between a smooth phase that we identify as the EW class and a rough phase with $\alpha^{3\rm{D}}_{\text{M}}\approx1$. We also show that, both in 1D and 3D, the small $\gamma$ limit can alternatively be thought as a short time limit $t\ll \gamma^{-1}$ within which the dynamics is described by the KPZ equation.

\paragraph{Continuous monitoring}

We begin by introducing the formalism of continuous monitoring. It is directly inspired from weak measurement and trajectory frameworks of quantum mechanics \citep{Aharanovweakmeasurement,DalibardQtrajectories,JacobsintroductiontoCmeasurement,BauerBernardTilloy_OpenQBrownianmotion,BernardJinShpielberg_2018} and can be thought as a simple hidden Markov process \cite{IntroductionHiddenMarkov}. 

In the absence of monitoring, the system undergoes a stochastic dynamics generated by ${\cal L}$
on a classical configuration space ${\cal M}({\cal C})$ with total number of configurations $\Omega$.
The time-evolution of the probability distribution ${\cal P}_{t}$
is given by the master equation
\begin{equation}
\label{eq:non_monitored_evolution}
\frac{d}{dt}{\cal P}_{t}({\cal C})={\cal L}({\cal P}_{t}({\cal C})).
\end{equation}
We assume that the stationary state is unique and is further given by the maximally
entropic state ${\cal P_{\infty}}=\Omega^{-1}$.
\begin{comment}
We now suppose that an external observer wishes to extract information
from the system by measuring it but in a non-totally efficient manner, for instance by taking blurry snapshots of it.\end{comment}
Weak monitoring takes place via an ancilla that couples to the system for a short
amount of time $\delta t$ such
that the generated correlation is of order $\delta t$ as well. 
Measuring the ancilla's state provides indirect and noisy information about the system which can be used to write a constrained stochastic evolution for ${\cal P}_t$. 
% Repeating this procedure with $M$ ancillae and taking
% the limit $\delta t\to0$, $M\to\infty$ while keeping $M\delta t:=t$
% fixed allows to express the continuous monitoring as a stochastic differential equation (SDE). 

Let $\Omega$ be the set $\{X_{1},\cdots, X_{N}\}:=\vec{X}$ where the $X_{j}$'s can take values $\pm1$: $+1$ corresponds to an occupied site while $-1$ to an empty one. 
We suppose that all sites will be independently
monitored. The ancilla monitoring site $j$ is also described
by a random variable $A_{j}$ which can take binary values $a_j\in\{-1,1\}$.
We denote by ${\cal P}(\vec{X},A_{j})$ the joint probability of the union system+ancillae
to be in a given configuration. We fix this probability distribution
to positively correlate the state of the system and of the ancilla :
\begin{equation}
\label{eq:ansatz_joint_distribution}
{\cal P}(\vec{X},A_{j})={\cal P}(\vec{X})\frac{1+\frac{\sqrt{\gamma\delta t}}{2}A_{j}X_{j}}{2}\;,
\end{equation}
where ${\cal P}(\vec{X})$ is the reduced probability of the system
only.
\begin{comment}
The interpretation of (\ref{eq:ansatz_joint_distribution}) is the following: the state of
the ancilla and the sites are positively correlated (e.g, if $A_{j}=1$,
it's more likely to find $X_{j}$ in $1$).
\end{comment}
Once a measurement of
the ancilla's state has been made with outcome $a_{j}$, the
probability distribution is updated with probability $1+\frac{\sqrt{\gamma\delta t}}{2}a_{j}\langle X_{j}\rangle$
to 
\begin{align}
{\cal P}(\vec{X}) & \to{\cal P}(\vec{X}|A_{j}=a_{j})={\cal P}(\vec{X})\frac{1+\frac{\sqrt{\gamma\delta t}}{2}a_{j}X_{j}}{1+\frac{\sqrt{\gamma\delta t}}{2}a_{j}\langle X_{j}\rangle},
\end{align}
where $\langle X_{j}\rangle:=\sum_{\{\vec{X}\}}X_{j}{\cal P}_{t}(\vec{X})$.
In the SM \cite{SM}, we show that repeating this procedure $M$ times and taking
the limit $M\to\infty$, $\delta t\to0$ while keeping $M\delta t=t$
fixed leads, in the It\=o prescription, to the following evolution for the probability distribution 
\begin{align}
\label{eq:time_evolution_monitoring}
d{\cal P}_{t}(\vec{X}) & =\frac{\sqrt{\gamma}}{2}{\cal P}_{t}(\vec{X})(X_{j}-\langle X_{j}\rangle_{t})dB_{t}^{j}\;,
\end{align}
where $dB_{t}^{j}$ are site-independent Brownian processes with variance
$dt$ and It\=o rules $dB_{t}^{j}dB_{t}^{k}=\delta_{j,k}dt$. Note that ${\cal P}_{t}$ is both a probability distribution
and a stochastic variable. Consequently, there
are two types of averages in the problem: $\langle\rangle$ denotes
average with respect to ${\cal P}_{t}$, while $\mathbb{E}[]$ denotes average with respect
to the Brownian processes $\{B_{t}^{j}\}$. 

As measurements occur independently on every site,
we obtain the stochastic evolution of the monitored system as the sum of (\ref{eq:non_monitored_evolution}) and (\ref{eq:time_evolution_monitoring}):
\begin{equation}
d{\cal P}_{t}={\cal L}({\cal P}_{t})dt+\frac{\sqrt{\gamma}}{2}\sum_{j}{\cal P}_{t}(\vec{X})(X_{j}-\langle X_{j}\rangle_{t})dB_{t}^{j}.\label{eq:generaleqwithmonitor}
\end{equation}
Note that, since ${\cal L}$ preserves the total probability and
$\sum_{\{\vec{X}\}}{\cal P}_{t}(\vec{X})(X_{j}-\langle X_{j}\rangle_{t})=0$,
the probability distribution $\mathcal{P}_t$ remains normalized at every time $t$ for each realization of
the process. 
\begin{comment}
In addition, continuous monitoring can not affect
the expectation of any physical observable $\mathbb{E}[\langle O(\vec{X})\rangle]$ since this quantity is linear in ${\cal P}_{t}$ and $d\mathbb{E}[{\cal P}_{t}]=\mathbb{E}[{\cal L}({\cal P}_{t})]dt$. This is expected since, contrary to the quantum case, classical measurements can't
affect the actual physical state of the system. 
Note, however, that quantities depending non-linearly
on ${\cal P}_{t}$ such as the Gibbs-Shannon entropy will
be affected by monitoring, even after the average $\mathbb{E}[]$
has been taken.\end{comment} 

\paragraph{Single-particle problem}

We now consider the specific case of a single random walker. For lightness, the following discussion will be for a 1D system of $N$ sites with periodic boundary conditions but generalization to higher dimensions is straightforward.
Let $p_{j}(t)$ be the probability for the particle to be at site
$j$ at time $t$. We choose ${\cal L}$ to be the discrete Laplacian weighted by a diffusion constant $D$, i.e ${\cal L}=D\Delta$ with $\Delta p_{j}:=p_{j-1}-2p_{j}+p_{j+1}$.
Starting from (\ref{eq:generaleqwithmonitor}), the evolution of $p_j$ in the presence of monitoring is given by
\begin{equation}
dp_{j}=D\Delta p_{j}dt+\sqrt{\gamma}p_{j}dW_{t}^{j}\;,\label{eq:singleparticlemonitoring}
\end{equation}
with $dW_{t}^{j}:=dB_{t}^{j}-\sum_{m}p_{m}dB_{t}^{m}$ (see \cite{SM} for the details of the calculation). 
Note that $dW_{t}^{j}$ are site-correlated Gaussian noises such that $\mathbb{E}[dW_{t}^{j}]=0$ and $\mathbb{E}[dW_{t}^{j}dW_{t}^{k}]=(\delta_{j,k}-(p_{j}+p_{k})+\sum_{m}p_{m}^{2})dt$.

\begin{comment}The stationary states corresponding to the first and second term
on the right hand side of Eq.(\ref{eq:singleparticlemonitoring}) are of quite different nature.
\end{comment}
The diffusive term favors the flat, maximally entropic distribution
$p_{j}=1/N$ while the measurement
term favors the $N$ pointer states $p_{j}=\delta_{j,k}\text{ for fixed }k\in\llbracket1,N\rrbracket$.
For finite $D$ and $\gamma$, the stationary distribution of this
model is non-trivial and, to the best of our knowledge, not known
with a notable exception for $N=2$. 
In the latter case, it turns
out that the dynamics is equivalent to the one of a single qubit undergoing both thermal
relaxation and quantum measurements and was treated in \citep{BauerBernardQubitThermal,BauerBernardTilloySpikes}.

Eq.(\ref{eq:singleparticlemonitoring}) is reminiscent of the stochastic
heat equation (SHE) with multiplicative noise \citep{BertiniSHEFeynmanKac}
except that the noise $dW^j_t$ is the sum of  a Brownian process and a non-local contribution $\sum_m p_m dB_t^m$.
\begin{comment}
$dW_{t}^{j}$
has average $0$ and correlations $dW_{t}^{j}dW_{t}^{k}=(\delta_{j,k}-(p_{j}+p_{k})+\sum_{m}p_{m}^{2})dt.$
\end{comment}
Nonetheless, it turns out that there is a formal correspondence between \eqref{eq:singleparticlemonitoring} and the SHE in the regime of small $\gamma$.

\begin{figure*}
\includegraphics[width=\linewidth]{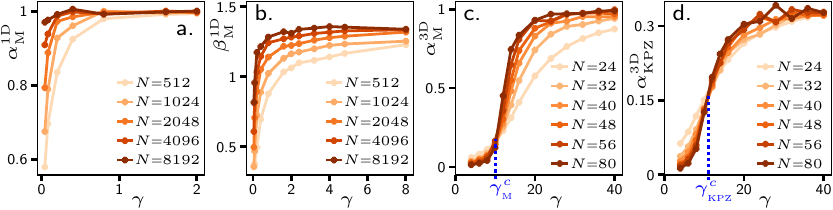}
\caption{\textbf{(a)}: Roughening exponent $\alpha^{\rm{1D}}_{\rm{M}}$ as a function of $\gamma$ for different system sizes.
\textbf{(b)}: Second growth exponent $\beta^{\rm{1D}}_{\rm{M}}$ as a function of $\gamma$ for different system sizes.
\textbf{(c)}: Roughening exponent $\alpha^{\rm{3D}}_{\rm{M}}$ as a function of $\gamma$ for different system sizes.
\textbf{(d)}: Roughening exponent $\alpha^{\rm{3D}}_{\rm{KPZ}}$ as a function of $\gamma$ for different system sizes.
Details about the methods used to extract the $\alpha$'s and $\beta$'s are in the SM \cite{SM}.
Note that, due to numerical limitations, only the roughening exponent was computed for the 3D case.
The 1D case \textbf{(a)}-\textbf{(b)} shows a size-dependent crossover from the KPZ exponents $\alpha^{\rm{1D}}_{\text{KPZ}}=1/2$, $\beta^{\rm{1D}}_{\text{KPZ}}=1/3$ to a new phase with exponents $\alpha^{1\rm{D}}_{\text{M}}\approx1$, $\beta^{1\rm{D}}_{\text{M}}\approx1.4$.
For the 3D case \textbf{(c)}, we observe that $\alpha^{\rm{3D}}_{\rm{M}}$ remains constant close to $0$ on a finite interval before jumping to $\alpha^{3\rm{D}}_{\text{M}}\approx1$ when $\gamma$ is greater than a critical value $\gamma^c_{\rm{M}}\approx 10$.
This step-like behavior indicates a phase transition from the EW class to a new universality class in 3D. 
For comparison, \textbf{(d)} shows the behavior of $\alpha^{\rm{3D}}_{\rm{KPZ}}$ as a function of $\gamma$ for the standard SHE \eqref{eq:discrete_SHE} in 3D where we also find $\gamma^{c}_{\rm{KPZ}}\approx 10$.}
\begin{comment}
{\bf Parameters:} $dt=0.001$, $D=1$. 
To obtain the width $w$, we averaged over at least $1000$ realizations of \eqref{eq:singleparticlemonitoring} such that the error bars of $w$ (corresponding to its standard deviation) are too small to be noticed on the plots.
\end{comment}
\label{fig:exponents}
\end{figure*}

\paragraph{Small $\gamma$ regime}
To highlight this correspondence, we now perform a perturbative analysis around $\gamma=0$ of \eqref{eq:singleparticlemonitoring} in the infinite system size limit $N\to\infty$. Suppose $p$ admits the small $\gamma$ expansion
\begin{equation}
\label{eq:expansion_small_gammma}
p=p^{(0)}+\sqrt{\gamma}p^{(1)}+\gamma p^{(2)}+\cdots\;,
\end{equation}
where $p^{(0)}$ is the stationary flat profile of the maximally
entropic state, i.e $p_{j}^{(0)}(t)=1/N,\forall(j,t)$.
Inserting (\ref{eq:expansion_small_gammma}) into (\ref{eq:singleparticlemonitoring}), we obtain the evolution of $p^{(1)}$ as
\begin{equation}
dp_{j}^{(1)}=D\Delta p_{j}^{(1)}dt+\frac{1}{N}(dB_{t}^{j}-\sum_{m}\frac{1}{N}dB_{t}^{m}).
\end{equation}
The term $\sum_{m}\frac{dB_{t}^{m}}{N}$ has mean $0$ and variance
$1/N$ so it is subleading in the limit $N\to\infty$. In this regime, we get 
\begin{equation}
dp_{j}^{(1)}\approx D\Delta p_{j}^{(1)}dt+p_{j}^{(0)}dB_{t}^{j}.\label{eq:order1}
\end{equation}
%As $p_{j}^{(0)}=\text{constant}$, \eqref{eq:order1} is nothing but the Edwards-Wilkinson
%\citep{EdwardsWilkinson} equation. 
The evolution of $p^{(2)}$ is obtained in a similar way:

\begin{align}
  dp_{j}^{(2)}=&D\Delta p_{j}^{(2)}dt+\\
  &\nonumber p_{j}^{(1)}(dB_{t}^{j}-\underbrace{\sum_{m}\frac{1}{N}dB_{t}^{m}}_{:={\rm I}})\underbrace{-\frac{1}{N}\sum_{m}p_{m}^{(1)}dB_{t}^{m}}_{:={\rm II}}.
\end{align}
As explained above, the variance of I scales as $1/N$. The variance
of II is given by $\mathbb{E}[\frac{1}{N^{2}}\sum_{j}(p_{j}^{(1)})^{2}]$.
Using translational invariance, we have on the other hand that $\mathbb{E}[(p_{j}^{(1)})^{2}]=\mathbb{E}[\frac{1}{N}\sum_{j}(p_{j}^{(1)})^{2}]$
so there is a factor of $N$ between the variance of the multiplicative
noise term $p_{j}^{(1)}dB_{t}^{j}$ and II. Thus, in the limit of
large $N$, we can neglect I and II to obtain

\begin{equation}
dp_j^{(2)}\approx D\Delta p_{j}^{(2)}dt+p_{j}^{(1)}dB_{t}^{j}.\label{eq:order2}
\end{equation}
This equation is structurally equivalent to Eq.(\ref{eq:order1}).
Thus, to order $\gamma$, the discrete SHE with multiplicative
noise
\begin{equation}
\label{eq:discrete_SHE}
dp_{j}=D\Delta p_{j}dt+\sqrt{\gamma}p_{j}dB_{t}^{j}
\end{equation}
is a good approximation of (\ref{eq:singleparticlemonitoring}).
\begin{comment}
\begin{equation}
\label{eq:mapping}
    D\Delta_{{\rm d}}p_{j}dt+\sqrt{\gamma}p_{j}dW_{t}^{j}=D\Delta_{{\rm d}}p_{j}+\sqrt{\gamma}p_{j}dB_{t}^{j}+\mathcal{O}(\gamma)\;.
\end{equation}
\end{comment}
Furthermore, the probability $p_j$ of the SHE is connected to the height $h_j$ of the KPZ equation via the Cole-Hopf transformation \citep{KPZpaper} $h_{j}:=\frac{1}{\sqrt{\gamma}}\log p_{j}$.
Indeed, using standard It\=o calculus on \eqref{eq:discrete_SHE}, we readily obtain the stochastic dynamics of $h_{j}$ as a discretized version of the celebrated KPZ equation (up to a linear shift in time $h_j \to h_j+\sqrt{\gamma}t$):
\begin{equation}
dh_{j}=(D\Delta h_{j}+D\sqrt{\gamma}\left(\nabla h_{j}\right)^{2}-\sqrt{\gamma})dt+dB_{t}^{j},\label{eq:discreteKPZmeas-1}
\end{equation}
where $\nabla$ is the discrete derivative $\nabla h_j := h_{j+1}-h_j$. 
Note that since $p_{j}\in[0,1]$, $h_{j}\in]-\infty,0]$.
Through its connection to the SHE, and therefore to the KPZ equation, we expect the dynamics of the monitored random walker to share common features with the physics of interface growth.
% Performing a similar Cole-Hopf transformation in \eqref{eq:singleparticlemonitoring} instead yields
% \begin{equation}
% dh_{j}=D\Delta_{{\rm d}}h_{j}+D\sqrt{\gamma}(\nabla_{{\rm d}}h_{j})^{2}-\sqrt{\gamma}(dW_{t}^{j})^{2}+dW_{t}^{j}+O(a^{3})\;,\label{eq:discreteKPZmeas}
% \end{equation}
% where $\nabla_{{\rm d}}h_{j}:=h_{j+1}-h_{j}$\textcolor{red}{1/a?}, and $a$ is the lattice spacing.
%Due to the mapping \eqref{eq:mapping}, 
One of the interesting quantities arising in the study of such interfaces is the so-called width $w$ defined as 
\begin{equation}
w:=(\frac{1}{N}\sum_{j}(h_{j}-\bar{h})^{2})^{1/2}\;,
\end{equation}
where $\bar{h}:=\frac{1}{N}\sum_{j}h_{j}$.
Starting from a flat initial profile, the Family-Vicsek (F-V) scaling relation \citep{FamilyVicsek,FractalconceptsBarbarasi}
conjectures that, for scale-invariant interfaces, the width should
behave as 
\begin{equation}
\label{eq:family_vicsek}
w\propto N^{\alpha}f\left(\frac{t}{N^{\alpha/\beta}}\right)
\end{equation}
with $f(u)\propto u^{\beta}$ for $u\ll1$ and $f(u)\propto{\rm const}$ for $u\gg1$.
The parameters $\alpha$ and $\beta$ are respectively called the roughening and growth exponents. 
For models within the KPZ universality class, it has been shown in 1D \cite{KPZpaper} that $\alpha^{\rm{1D}}_{\rm{KPZ}}=1/2$ and $\beta^{\rm{1D}}_{\rm{KPZ}}=1/3$.
We thus expect that the width of the log-probability of the monitored random walker will follow \eqref{eq:family_vicsek} with KPZ exponents when $\gamma$ is small (see Fig.\ref{fig:exponents}-a and Fig.\ref{fig:exponents}-b).

Importantly, one can alternatively think of the small $\gamma$ expansion as a short time limit. Indeed, at short times, $t\ll\gamma^{-1}$, the probability profile will be close to the initial flat distribution. If we assume that the leading term in $p_j$ scales like $1/N$, it is easy to check that $\mathbb{E}[(\sum_m p_m dB_t^m)^2] \approx O(N^{-1})dt$ so that the contribution of the non-local part of $dW_t^j$ is subleading. 

However, in the long-time regime $t\gg \gamma^{-1}$, we expect to be pushed out of the KPZ regime as the roughening of the probability profile makes the non-local term of the noise grow.

In addition, the mapping to KPZ physics at short times and/or small $\gamma$ tells us that a roughening phase transition from a smooth to a rough interface should occur in 3D and above \citep{NumericsPhaseTransition_MOSER1991215,PhaseTransition_HongKesslerSander,NumericsPhasetransition_Torres_2018,nonperturbativeRGLeonieCanet}. Indeed, at small $\gamma$, we can neglect the contribution of the non-local part of the noise and thus the perturbative dynamic renormalization flow leads to similar flow equations than those of the KPZ equations \citep{KPZpaper}. In the smooth phase, the roughening term becomes irrelevant so we can safely neglect the non-local part of the noise. There, we expect that our model will flow to the same universality class as the KPZ equation, i.e the Edwards-Wilkinson (EW) class. However, this similarity should break down in the roughening phase where we expect \eqref{eq:singleparticlemonitoring} to flow to a different universality class than KPZ.

Although the analytical investigation of the strong $\gamma$ regime is beyond the scope of this paper, we performed a series of numerical simulations of \eqref{eq:singleparticlemonitoring} in 1D and 3D to confirm the previous qualitative reasoning regarding the renormalization flow.
% \begin{equation}
% dh_{j}=D\Delta_{{\rm d}}h_{j}+D\sqrt{\gamma}(\nabla_{{\rm d}}h_{j})^{2}-\sqrt{\gamma}(dW_{t}^{j})^{2}+dW_{t}^{j}+O(a^{3})\;,\label{eq:discreteKPZmeas}
% \end{equation}
% where $\nabla_{{\rm d}}h_{j}:=h_{j+1}-h_{j}$, and $a$ is the lattice spacing.
% Again, we see that if the $dW_{t}^{j}$'s were white noises, \ref{eq:discreteKPZmeas} would
% just be a discretization of the KPZ equation. Note however that the
% $dW_{t}$ terms explicitly break 
% the translation invariance along the growth direction $h\to h+\delta h$
% of the KPZ equation.

\paragraph{Numerical results} We started all our simulations with a flat initial profile
$p_{j}(t=0)=1/N^d$, i.e $h_{j}(t=0)=-\frac{d}{\sqrt{\gamma}}\log N$ with $d$ being the dimension.
% ,which corresponds to the maximally entropic state. 
We simulated \eqref{eq:singleparticlemonitoring} using a standard Euler-Maruyama scheme and took the logarithm for
every single realization to obtain the evolution of the process
$h_{j}$. Details about the numerical methods, convergence check and finite-size scaling are provided in the SM \cite{SM}.

\begin{figure*}
\includegraphics[width=\linewidth]{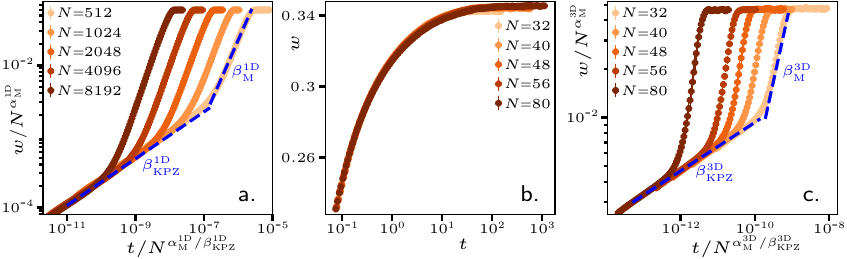} 
\caption{\textbf{(a)}: Log-Log plot of the rescaled width $w/N^{\alpha^{1\rm{D}}_{\rm{M}}}$ as a function of the rescaled time $t/N^{\alpha^{1\rm{D}}_{\rm{M}}/\beta^{1\rm{D}}_{\rm{KPZ}}}$ for a measurement rate $\gamma=4$ in 1D. Exponents: $\alpha^{1\rm{D}}_{\rm{M}}=1$ and $\beta^{1\rm{D}}_{\rm{KPZ}}=0.34$.
\textbf{Parameters}: $D=1$, $dt=0.001$.
% In the short time regime, all curves collapse onto a single one characterized by the growth exponent $\beta^{1\rm{D}}_{\rm{KPZ}}$. 
\textbf{(c)}: Log-Log plot of the rescaled width $w/N^{\alpha^{3\rm{D}}_{\rm{M}}}$ as a function of the rescaled time $t/N^{\alpha^{3\rm{D}}_{\rm{M}}/\beta^{3D}_{\rm{KPZ}}}$ for a measurement rate $\gamma=36$ in 3D. Exponents: $\alpha^{3\rm{D}}_{\rm{M}}=1$ and $\beta^{3D}_{\rm{KPZ}}=0.15$.
\textbf{Parameters}: $D=1$, $dt=0.0002$.
% In the short time regime, all curves collapse onto a single one characterized by the KPZ growth exponent $\beta^{3D}_{\rm{KPZ}}$.
\textbf{(a) and (c)}: In accordance with the perturbative analysis, which is valid at short times, the initial growth exponent is always KPZ-like.
At intermediate and large $\hat{t}$ however, the F-V scaling of the width flows to a new universality class characterized by $\beta^{\rm{1D}}_{\rm{M}}\simeq 1.4$ and $\alpha^{\rm{1D}}_{\rm{M}}\approx1$ in 1D or $\beta^{\rm{3D}}_{\rm{M}}\simeq 1.2$ and $\alpha^{\rm{3D}}_{\rm{M}}\approx1$ in 3D.
\textbf{(b)}: Linear-Log plot of the width $w$ as a function of time $t$ for a measurement rate $\gamma=4$ in 3D. 
As $\gamma<\gamma^{c}_{\rm{M}}$, the system is in the smooth EW phase and the width do not show any dependency on the system size.
\textbf{Parameters}: $D=1$, $dt=0.002$.
}
\label{fig:FV}
\end{figure*}

We plot on Fig.\ref{fig:FV}-a the rescaled width $\hat{w}=w/N^{\alpha^{\rm{1D}}_{\rm{M}}}$ as a function of the rescaled time $\hat{t}=t/N^{\alpha^{\rm{1D}}_{\rm{M}}/\beta^{\rm{1D}}_{\rm{KPZ}}}$ in 1D for different system sizes when $\gamma=4.0$. 
In agreement with the connection to KPZ at short times, all curves collapse on the power law $t^{\beta^{\rm{1D}}_{\rm{KPZ}}}$ at small $\hat{t}$.
However, beyond this regime, the F-V scaling of the width flows to a new universality class characterized by $\beta^{\rm{1D}}_{\rm{M}}\approx 1.4$ and $\alpha^{\rm{1D}}_{\rm{M}}\approx1$.

Fig.\ref{fig:FV}-c is a similar plot but performed in 3D when $\gamma=36$ and for which the rescaled width and time are respectively given by $\hat{w}=w/N^{\alpha^{\rm{3D}}_{\rm{M}}}$ and $\hat{t}=t/N^{\alpha^{\rm{3D}}_{\rm{M}}/\beta^{\rm{3D}}_{\rm{KPZ}}}$.
At small $\hat{t}$, all curves collapse on the expected power law $t^{\beta^{\rm{3D}}_{\rm{KPZ}}}$ while beyond this regime the F-V scaling flows to a new universality class characterized by $\beta^{\rm{3D}}_{\rm{M}}\approx 1.2$ and $\alpha^{\rm{3D}}_{\rm{M}}\approx1$.

\begin{comment}
As reported in e.g \cite{Dasgupta_NumericsKPZ,Lam_AnomalyNumericsKPZ}, such deviation from the F-V scaling could be attributed to the fact that a strong non-linearity is necessary to drive \eqref{eq:discreteKPZmeas-1} to the KPZ fixed point.
\end{comment}
Finally, on Fig.\ref{fig:FV}-b, we plot $w$ as a function of $t$ in 3D for different system sizes when $\gamma=4$.
For this value of $\gamma$, the KPZ equation flows toward the smooth EW class where we expect the non-local part of the noise to be irrelevant.
In agreement with this intuition, Fig.\ref{fig:FV}-b shows indeed that the width does not scale with $N$.
\begin{comment}
We plot on  Fig.\ref{fig:exponents}-a the dependence of these exponents with $\gamma$. We interprete the data as a crossover between the KPZ class at short times/small $\gamma$ and a new universality class characterized by exponents $\alpha^{1\rm{D}}_{\text{M}}\approx 1$ and $\beta^{1\rm{D}}_{\text{M}}\approx 1.4$ at long times/strong $\gamma$.
\end{comment}

We report on Fig.\ref{fig:exponents} the critical exponents as a function of $\gamma$ for simulations performed on $1$D and $3$D lattices.
For the former case (Fig.\ref{fig:exponents}-a and Fig.\ref{fig:exponents}-b), we observe a size-dependent crossover between the KPZ phase and a new phase characterized by exponents $\alpha^{1\rm{D}}_{\text{M}}\approx1$ and $\beta^{1\rm{D}}_{\text{M}}\approx1.4$. 
For the 3D case, we report on Fig.\ref{fig:exponents}-c the existence of a finite range over which $\alpha^{\rm{3D}}_{\rm{M}}$ is close to $0$, thereby indicating the presence of two distinct phases separated by a critical value $\gamma^{c}_{\rm{M}}\approx 10$. 
For comparison, \ref{fig:exponents}-d shows the behavior of $\alpha^{\rm{3D}}_{\rm{KPZ}}$ with respect to $\gamma$ when simulating the SHE equation \eqref{eq:discrete_SHE} in 3D where we find $\gamma_{\text{KPZ}}^c\approx10$. The fact that the two critical values for the SHE and our model are close corroborate our previous qualitative reasoning concerning the smooth phase in 3D.
As we are only interested in the existence of a MIPT, we only reported the behavior of $\alpha^{\rm{3D}}_{\rm{M}}$ as the systematic determination of $\beta^{\rm{3D}}_{\rm{M}}$ is more involved and left for future works.

\begin{comment}
The data available leads us to conjecture that the condition for the emergence of this new regime is that the transition time $\tilde{t}$ must be smaller than the typical saturation time $t^*\approx N^{\alpha/\beta}$.
When this second growth regime is absent from the simulations, there are two possibilities: either the system size $N$ considered is too small or $\tilde{t}\to\infty$, which would be an indication of a measurement-induced transition. 
% Since simulations are always performed at a fixed value of $N$, whenever the second growth regime doesn't appear, there is either the possibility that the system size considered is too small or that $\tilde{t}\to\infty$ which would be an indication of a measurement-induced transition. 
Distinguishing between these two possibilities would require a careful analysis and extrapolation at $N\to\infty$ of the behavior of $\tilde{t}$ as a function of $\gamma$ which is beyond the scope of this work and we defer it to subsequent studies. DANS L'APPENDICE?
\end{comment}
\paragraph{Conclusion and perspectives}
In this paper we introduced and studied a model for a single
random walker undergoing continuous measurement. 
In the regime of weak monitoring, we mapped the time evolution of its probability distribution onto the SHE.
We deduced that, in this regime, the
width of the log probability follows the F-V scaling relation of the KPZ universality class. 
In 1D, this corresponds to roughening and growth exponents  $\alpha^{\rm{1D}}_{\rm{KPZ}}=1/2$ and $\beta^{\rm{1D}}_{\rm{KPZ}}=1/3$. Beyond weak monitoring, we numerically find in 1D that increasing $\gamma$ leads to a crossover from the KPZ class to a new universality class with exponents $\alpha^{1\rm{D}}_{\text{M}}\approx1$ and $\beta^{1\rm{D}}_{\text{M}}\approx 1.4$. In 3D, we showed, again numerically, that this crossover becomes a phase transition between a smooth phase that we identify as the EW class and a new phase with $\alpha^{3\rm{D}}_{\text{M}}\approx1$.

Our study is one of the first characterization of a MIPT in classically monitored systems and opens the door to several interesting questions. It would be most desirable to have a better analytical characterization of the strong $\gamma$ regime. Since perturbative methods ought to fail there, non-perturbative RG methods such as the one presented in \citep{nonperturbativeRGLeonieCanet} may be employed there. 

While we only considered a flat profile, it is known
that different initial distributions leads to different
universality classes in KPZ physics \citep{ReviewKPZ}.
Thus, it would be interesting to investigate various initial states such as wedge or Brownian conditions to assess the effect of continuous monitoring on their corresponding exponents.

Finally, while we only studied a single particle, the continuous
measurement process \eqref{eq:generaleqwithmonitor} is easily generalized to more intricate, many-body
interacting problems. A natural extension would be to consider the symmetric
simple exclusion process (SSEP), which describes multiple diffusive
particle with hard-core repulsion. Interestingly the SSEP can be promoted
to a quantum version called the QSSEP \citep{BauerBernardJin_EquilibriumQSSEP,JinQSSEP}.
% which can be thought of as a continuous-time version of random
% unitary circuits \cite{EntanglementRUCNahum}. 
The study of both SSEP and QSSEP would thus provide a unified framework to disentangle the properties specific to quantum and classical systems under continuous monitoring.

\begin{acknowledgments}
\textbf{Acknowledgements} T.J and D.M thanks D. Bernard, N. Caballero, L. Canet, A. Krajenbrink, V. Lecomte, P. Ledoussal,
M. Medenjak, C. Nardini and L. Piroli for useful discussions. We are grateful to Y. Sato for his game TimeBomb$^\copyright$  which served as an inspiration for this work. T.J acknowledges support from the Swiss National Science Foundation under Division II.
\textit{During the writing of the manuscript, it came to our knowledge that two works with a similar objective of studying measurement effects on chaotic, classical systems but with a focus on phase transition were put as preprints \cite{MeasurementsystemclassicWillsher, Knolle_Bridgingthegap}}
\end{acknowledgments}

\bibliographystyle{apsrev4-2}
\bibliography{BiblioRWmeasurement}

\onecolumngrid\pagebreak\newpage{}

\appendix

\section*{Supplementary material}

\subsection{Derivation of the model}
\label{sec:Weak-measurements}

In this supplementary material, we present the derivation of \eqref{eq:generaleqwithmonitor} and \eqref{eq:singleparticlemonitoring}
of the main text.

The state of system is described by a set of random variables $\{X_{1},\cdots, X_{N}\}:=\vec{X}$
which can take values $\pm1$ where $+1$ corresponds to an occupied
site and $-1$ to an empty site. 
The ancilla monitoring site $j$
is also described by a random variable $A_{j}$ which can take binary
values $\{-1,1\}$. 
% In what follows we will be a bit sloppy with notation
% and denote the random variable and the value they are evaluated to
% indifferently. 
We denote by ${\cal P}(\vec{X},A_{j})$ the joint probability
of the union system+ancilla to be in a given configuration. 
We fix this probability distribution to
\begin{equation}
\label{eq:joint_distribution_appendix}
{\cal P}(\vec{X},A_{j})={\cal P}(\vec{X})\frac{1+\varepsilon A_{j}X_{j}}{2}\;,
\end{equation}
with $\varepsilon=\frac{\sqrt{\gamma\delta t}}{2}$ being a small paramater
and ${\cal P}(\vec{X})$ being the reduced probability of the system only.
The joint distribution \eqref{eq:joint_distribution_appendix} implies that the state of ancilla $j$ and its corresponding site are positively correlated: if $A_{j}=1$, it's
more likely to find $X_{j}$ in state $1$. 
The reduced probability ${\cal P}(A_{j})$ for the $j$-th ancilla is given by 
\begin{align}
{\cal P}(A_{j}) =\sum_{\{\vec{X\}}}{\cal P}(\vec{X},A_{j})
=\frac{1}{2}(1+\varepsilon A_{j}\langle X_{j}\rangle).
\label{eq:reduced_proba_appendix}
\end{align}
Once a measurement of the state of the ancilla has been made with
outcome $A_{j}$, the probability distribution is updated with probability
${\cal P}(A_{j})$ to 
\begin{align}
\label{eq:projection_appendix}
{\cal P}(\vec{X}) & \to{\cal P}(\vec{X}|A_{j})={\cal P}(\vec{X})\frac{1+\varepsilon A_{j}X_{j}}{1+\varepsilon A_{j}\langle X_{j}\rangle}\;,
\end{align}
which is just Bayes law. 
We now repeat this procedure
$M$ times with a fresh ancilla $A_{j}^{(n)}$ indexed by $n\in\llbracket 1,M\rrbracket$. 
Expanding \eqref{eq:projection_appendix} to order $\varepsilon^{2}$, we get: 
\begin{align}
{\cal P}_{n+1}(\vec{X}|A_{j}^{(n)}) & ={\cal P}_{n}(\vec{X})\left(1+\varepsilon A_{j}^{(n)}X_{j}\right)\left(1-\varepsilon A_{j}^{(n)}\langle X_{j}\rangle_{n}+\varepsilon^{2}\left(A_{j}^{(n)}\langle X_{j}\rangle_{n}\right)^{2}\right)+O(\varepsilon^{3})\\
 & ={\cal P}_{n}(\vec{X})\left(1+\varepsilon A_{j}^{(n)}\left(X_{j}-\langle X_{j}\rangle_{n}\right)-\varepsilon^{2}\left(X_{j}\langle X_{j}\rangle_{n}-\langle X_{j}\rangle_{n}^{2}\right)\right)+O(\varepsilon^{3})\;,
 \label{eq:time_evolution_appendix}
\end{align}
where $\langle\rangle_{n}$ indicates that the average has to be taken
with ${\cal P}_{n}$ and we used that $(A_{j}^{(n)})^{2}=1$. 
The signal is defined as the sum of the measurement outputs on the ancilla, i.e $S_{j,M}:=\sum_{n=1}^{M}\varepsilon A_{j}^{(n)}$, from which we deduce its increment
$S_{j,M+1}-S_{j,M}=\varepsilon A_{j}^{(M+1)}$. 
Using \eqref{eq:reduced_proba_appendix}, we further obtain that
\begin{equation}
\langle\varepsilon A_{j}^{(n)}\rangle  =\frac{\varepsilon}{2}(1+\varepsilon\langle X_{j}\rangle_{n})-\frac{\varepsilon}{2}(1-\varepsilon\langle X_{j}\rangle_{n})=\varepsilon^{2}\langle X_{j}\rangle_{n}\;,\qquad\qquad
\langle(\varepsilon A_{j}^{(n)})^{2}\rangle =\varepsilon^{2}\;.
\end{equation}
These relations show that in the limit $\varepsilon\to0$, $M\delta t=t$,
the signal converges in law towards a process described by the following
stochastic differential equation
\begin{equation}
dS_{j,t}=\frac{\gamma}{4}\langle X_{j}\rangle_{t}dt+\frac{\sqrt{\gamma}}{2}dB_{t}^{j}\;,
\end{equation}
where $j$ is the site index and $B_{t}^{j}$ is a $0$-mean, site-independent Brownian process of variance $\mathbb{E}[B_{t}^{j}B_{t}^{i}]=\delta_{ij}dt$.
We can now replace the $A_{j}$'s in the evolution equation for
the probability \eqref{eq:time_evolution_appendix} to get 
% \begin{align}
% d{\cal P}_{t}(\vec{X}) & ={\cal P}_{t}(\vec{X})\left(\left(\frac{\gamma}{4}\langle X_{j}\rangle_{t}dt+\frac{\sqrt{\gamma}}{2}dB_{t}^{j}\right)\left(X_{j}-\langle X_{j}\rangle_{t}\right)-\frac{\gamma}{4}\left(X_{j}\langle X_{j}\rangle_{t}-\langle X_{j}\rangle_{t}^{2}\right)dt\right),\nonumber \\
% d{\cal P}_{t}(\vec{X}) & =\frac{\sqrt{\gamma}}{2}{\cal P}_{t}(\vec{X})\left(X_{j}-\langle X_{j}\rangle_{t}\right)dB_{t}^{j}.
% \end{align}
\begin{equation}
d{\cal P}_{t}(\vec{X})={\cal P}_{t}(\vec{X})\left(\left(\frac{\gamma}{4}\langle X_{j}\rangle_{t}dt+\frac{\sqrt{\gamma}}{2}dB_{t}^{j}\right)\left(X_{j}-\langle X_{j}\rangle_{t}\right)-\frac{\gamma}{4}\left(X_{j}\langle X_{j}\rangle_{t}-\langle X_{j}\rangle_{t}^{2}\right)dt\right) =\frac{\sqrt{\gamma}}{2}{\cal P}_{t}(\vec{X})\left(X_{j}-\langle X_{j}\rangle_{t}\right)dB_{t}^{j}.
\end{equation}
If measurement processes occur independently on every site and
we include the internal stochastic dynamics ${\cal L}$ of the system, we obtain the following SDE for ${\cal P}_{t}$:
\begin{equation}
d{\cal P}_{t}={\cal L}({\cal P}_{t})dt+\frac{\sqrt{\gamma}}{2}\sum_{j}{\cal P}_{t}(\vec{X})(X_{j}-\langle X_{j}\rangle_{t})dB_{t}^{j}\;,\label{eq:generaleqwithmonitor-1}
\end{equation}
which is \eqref{eq:generaleqwithmonitor} in the main text.

We now specify to the particular case of a single random walker on
a discrete lattice of $N$ sites with periodic boundary conditions.
In this case, we define $p_{j}(t):={\cal P}_{t}(X_{1}=-1,\cdots,X_{j}=1,\cdots, X_{N}=-1)$ and we further have that
\begin{equation}
\label{eq:specific_walker}
{\cal L}(p_{j}) =D(p_{j-1}-2p_{j}+p_{j+1}), \qquad\qquad
\langle X_{j}\rangle_{t} =\sum_{j'\neq j}-p_{j'}+p_{j}=-1+2p_{j}\;.
\end{equation}
Inserting \eqref{eq:specific_walker} into \eqref{eq:generaleqwithmonitor-1}, we obtain the time-evolution of $p_j$ (Eq.\eqref{eq:singleparticlemonitoring} in main text) as
\begin{align}
dp_{j} & =D(p_{j-1}-2p_{j}+p_{j+1})+\frac{\sqrt{\gamma}}{2}\left(\sum_{m\neq j}p_{m}(-2p_{j})dB_{t}^{m}+p_{j}(2-2p_{j})dB_{t}^{j}\right),\nonumber \\
 & =D(p_{j-1}-2p_{j}+p_{j+1})+\sqrt{\gamma}p_{j}\left(dB_{t}^{j}-\sum_{m}p_{m}dB_{t}^{m}\right)\;.
\end{align}
% which is Eq.(\ref{eq:singleparticlemonitoring}) in the main text.

\subsection{Numerical methods}
\label{sec:numerical_methods}

\begin{figure}
\includegraphics[width=\linewidth]{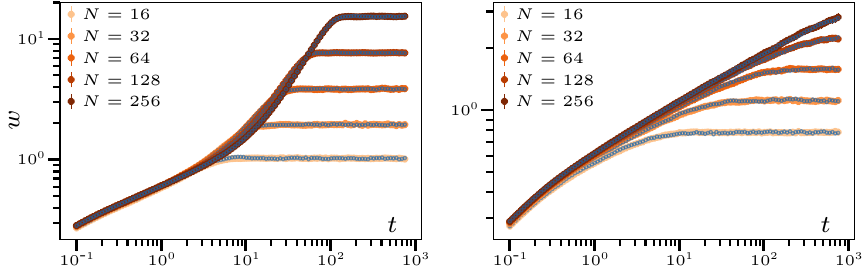} 
\caption{{\bf Left:} Width as a function of time for $dt=0.001$ (orange shaded dots) and for $dt=0.0005$ (unfilled blue circles) at different values of $N$ and fixed $\gamma=4$. {\bf Right:} Same plot at fixed $\gamma=0.02$. 
In both cases, the numerical algorithm has converged to the solution. {\bf Parameters:} $D=1$. To obtain the width $w$, we averaged over at least $1000$ realizations of \eqref{eq:singleparticlemonitoring} such that the error bars of $w$ (corresponding to its standard deviation) are too small to be noticed on the plots.}
\label{fig:check_numerics}
\end{figure}

We hereafter describe the methods employed to numerically integrate \eqref{eq:singleparticlemonitoring}.
We first note the peculiar structure of the noise term $dW^j_t$ in \eqref{eq:singleparticlemonitoring}: it is a multiplicative multi-dimensional noise.
Thus, \eqref{eq:singleparticlemonitoring} falls into the class of stochastic differential equations taking the form
\begin{equation}
    \label{eq:multidimensional_SDE}
    \frac{dx_j}{dt} = f_j(\{\bx\}) + \sum_{i=1}^{N} g_{ji}(\{\bx\})\xi_i
\end{equation}
where $\xi_i$'s are gaussian white noises such that $\langle\xi_i(s)\xi_k(s')\rangle=\delta_{ik}\delta(s-s')$, and $f_j$ and $g_{ji}$ are functions of the set of position $\{\bx\}=\{x_k,\ k\in [1,..,N]\}$.
Numerical integration schemes for SDEs of type \eqref{eq:multidimensional_SDE} have been discussed in \cite{mannella1997numerical} and chapter 7 of \cite{moss_experiments_1989}.
The combination of the multiplicative and multi-dimensional nature of the noise in \eqref{eq:multidimensional_SDE} renders usual higher order Runge-Kutta-based SDE algorithms inoperative. 
As described in \cite{moss_experiments_1989}, the two numerical schemes available for integrating \eqref{eq:multidimensional_SDE} are both of order $dt$ at maximum.
The first one is an Euler-Maruyama scheme, (ie simple forward Euler), which allows for a straightforward integration of \eqref{eq:multidimensional_SDE} in Ito prescription. 
The second one is a first order Runge-Kutta scheme with an approximate closure valid up to $dt$: it allows for numerical integration of \eqref{eq:multidimensional_SDE} directly in Stratonovitch prescription.
As we studied \eqref{eq:singleparticlemonitoring} and within Ito formalism in the main text, we naturally choose the former Euler-Maruyama algorithm to perform our numerical integrations.
To check the convergence of the algorithm, we divided the time step $dt$ by two and verified the stability of our results (see Fig~\ref{fig:check_numerics}).
We further constantly monitored the probabilities $p_j$'s and choose a sufficiently low time step $dt$ ensuring that $p_j(t)>0$ for all $j\in[1,..,N]$ at every time $t$.
Finally, we also monitored the conservation of probabilities and checked that $\sum_j p_j(t)=1$ at every time $t$.
\\
The width $w$ at fixed $\gamma$ and fixed system size $N$ was obtained by averaging over at least 1000 realizations of \eqref{eq:singleparticlemonitoring}.
We made sure that simulations ran long enough for $w$ to effectively reach its plateau value at large time.
To compute the roughening exponent $\alpha$ at fixed $\gamma$, we performed a linear fit of the width's plateau value $w(t=\infty)$ as a function of the system size $N$ in Log-Log.
From \eqref{eq:family_vicsek}, we indeed have that $\log(w(t=\infty))\sim\alpha \log(N)$: the coefficient of the later linear fit gives $\alpha$.
\\
To extract the growth exponent $\beta_{\rm{KPZ}}$, we performed a linear fit of the width $w$ against $t$ in Log-Log at small times. 
From our small-$\gamma$ perturbative analysis and \eqref{eq:family_vicsek}, we have that $\log(w(t))\sim \beta_{\rm{KPZ}} \log(t)$ at small times and the coefficient of the later linear fit thus gives $\beta_{\rm{KPZ}}$.
In practise, we made sure to apply the linear fit only for small times where the logarithm of the width increases linearly with respect to $t$.
Finally, to extract the second growth exponent $\beta_{\rm{M}}$, we performed a linear fit of the width $w$ against $t$ in Log-Log at intermediate times.
We made sure to perform the later fit in between the initial KPZ-like growth regime characterized by $\beta_{\rm{KPZ}}$ and the plateau regime characterized by $\alpha_{\rm{M}}$.

\end{document}